\documentclass[lineno]{jfm}
\usepackage{graphicx,epstopdf,epsfig,subfig,float}
\usepackage{amsmath,amssymb,amsfonts,gensymb,color}
\usepackage{mathrsfs,mathtools,bm,siunitx}
\usepackage{array,booktabs,multicol,multirow}
\usepackage{newtxtext,newtxmath}
\usepackage{natbib,hyperref,orcidlink,url}
\hypersetup{
	colorlinks = true,
	urlcolor   = blue,
	linkcolor  = blue,
	citecolor  = blue,
}

\newcommand{\RomanNumeralCaps}[1]
\linenumbers

\allowdisplaybreaks[4]
\title{Isotropic extension of first-order wave equations}
\author{Shengqi Zhang\corresp{\email{szhang@eitech.edu.cn}}}
\affiliation{College of Engineering, Eastern Institute of Technology, Ningbo 315201, P. R. China.}
\begin{document}
\maketitle
\begin{abstract}
The anisotropy of many one-dimensional and first-order-in-time (T$^1$) scalar wave equations (e.g., Korteweg-de Vries and Camassa-Holm) limits their physical completeness and applicability to bidirectional/high-dimensional systems. We define the T$^n\Lambda^m$ isotropic extension consisting of temporal order elevation and spatial tensorization, which is the only possible approach to eliminate anisotropy while preserving original solutions. Our analysis finds that the Burgers equation exhibits T$^{\mathbb{N}_+}\Lambda^{2\mathbb{N}+1}$ extensibility and the Korteweg-de Vries (KdV) equation exhibits the T$^{2\mathbb{N}_+}\Lambda^{2\mathbb{N}}$ extensibility. The T$^2\Lambda^0$ extension of the KdV equation leads to the corresponding isotropic T$^2$ equation (KdV$^2$) for shallow water dynamics, which is physically more complete and suitable for 2D generalization. In addition to inheriting all KdV solutions and conservation laws, the KdV$^2$ equation also provides linearly stable corrections to the Boussinesq equation. In contrast, the KdV-Burgers equation is inherently anisotropic as it fails to exhibit any T$^n\Lambda^m$ extensibility.
\end{abstract}
\section{Motivation for isotropic extension}
Spatial isotropy, as a basic symmetry in physics, is an important criterion for evaluating the physical completeness or broad applicability of mathematical models. However, various important wave equations are anisotropic. In many circumstances, wave equations derived from simplified analysis appear in one-dimensional (1D) and first-order-in-time (T$^1$) scalar forms. Although capable of successfully describing specific physical phenomena in hydrodynamics, turbulence, and condensed matter physics, many of these equations have their bidirectional or higher-dimensional applicability limited by anisotropy. 

For example, the classic KdV equation \citep{Korteweg1895Onthe}, with widespread attention due to its conservation properties and soliton solutions, only describes unidirectional shallow water waves. Evidently, it does not satisfy isotropy: to obtain solutions propagating in the opposite direction, one must reverse the sign of all terms containing spatial derivatives. This forbids the coexistence of left- and right-propagating wave solutions for a single equation. Consequently, such anisotropy also forbids simple generalization of the KdV equation to 3D horizontally homogeneous shallow water.

In contrast, the Boussinesq equation \citep{Boussinesq1871aTh,Boussinesq1871bTh} is isotropic and bidirectional in 1D and can be naturally generalized to 2D. However, despite certain similarities \citep{Miles1981TheKorteweg,Hong2024Onthe}, the solutions of the KdV equation do not strictly satisfy the Boussinesq equation. There is another 2D KdV-type equation derived for lattice systems based on a phenomenological model \citep{Tasi1980Asecond}, which is isotropic in a continuum limit. However, its continuum limit still cannot preserve the solutions of the KdV equation. Another 2D generalization based on the KdV equation is the Kadomtsev-Petviashvili equation \citep{Kadomtsev1970Onthe}, which characterizes the dispersion perpendicular to the wave propagation direction by adding a Laplacian term in the transverse direction, but it is anisotropic. In short, none of the existing equations similar to the KdV equation can simultaneously preserve the KdV solutions and satisfy spatial isotropy.

If an isotropic equation can be formulated based on the KdV equation, whose solutions include all well-behaved solutions of the KdV equation, it will provide important tools for fundamental theory and engineering applications. In particular, isotropic extension can greatly broaden the application scope of mathematical theories such as soliton theory towards bidirectional or higher-dimensional circumstances. With the expected isotropic extension, the classic unidirectional solitons can be generalized to describe waves in 3D homogeneous shallow water, with propagation directions freely chosen within the horizontal plane.

Similar situations may also exist for equations such as the KdV-Burgers \citep{Johnson1970Anon}, mKdV \citep{Miura1968Kortewegde}, mKdV-Burgers \citep{Wang1996Exactsolutions}, Camassa-Holm \citep{Camassa1993Anintegrable}, and Degasperis-Procesi \citep{Degasperis1999AsymptoticIntegrability} equations. Exploring the possibility and approaches to extend these equations into isotropic forms holds significant importance for enhancing physical understanding and promoting the applicability of existing theories.

Motivated by these considerations, the focus of this paper is to explore possible extensions of one-dimensional and first-order-in-time scalar wave equations, making them suitable for isotropic systems while preserving their original solutions.
\section{Isotropic extension framework}
For a real scalar field $\phi(x,t)\in\mathbb{R}$ in 1D space ($x\in\mathbb{R}$), we first consider the T$^1$ wave equation of the form:
\begin{equation}\label{eqn:original}
	\partial_t\phi=F[\phi]=\sum\limits_{p\in\mathbb{N},q\in\mathbb{N}_+}F_{p,q}[\phi],
\end{equation}
where $\mathbb{N}$ is the set of natural numbers including $0$, $\mathbb{N}_+$ is the set of positive integers, and $F_{p,q}$ is an operator constructed using only multiplication, addition, and local spatial differentiation. $F_{p,q}[\phi]$ contains $(p,q)$ terms that comprise spatial derivative $\partial_x$ of total order $p$ and $\phi$ of total power $q$. For example, $F_{2,3}[\phi]$ can contain $(2,3)$ terms such as $\partial_x^2(\phi^3)$, $\phi(\partial_x\phi)^2$, $\phi^2\partial_x^2\phi$, etc.

It is evident that only spatial or temporal differentiation on both sides of equation \eqref{eqn:original} can obtain new equations that preserve all original solutions. Since spatial differentiation cannot reverse anisotropy, temporal differentiation is the only possible analytical operation for solution-preserving isotropic extension.

We define the (temporal order) elevation as taking the time derivative of both sides of the T$^{n-1}$ equation, and using the original equation \eqref{eqn:original} to eliminate terms containing the first-order time derivative, yielding the T$^{n}$ equation:
\begin{subequations}\label{eqn:OrderElev}\begin{gather}
	\label{eqn:OrderElevA}
	\partial_t^{n}\phi
	=F^{(n)}[\phi]=\sum\limits_{p\in\mathbb{N},q\in\mathbb{N}_+}F^{(n)}_{p,q}[\phi],
	\\
	\label{eqn:OrderElevB}
	n\geq2:~
	F^{(n)}_{p,q}[\phi]
	=\sum\limits_{\substack{p'+p''=p\\q'+q''=q+1}}F'^{(n-1)}_{p',q'}\left[\phi,F_{p'',q''}[\phi]\right],
\end{gather}\end{subequations}
where $F^{(1)}=F$. For an arbitrary operator $G$ of $\phi$, $G'$ is an operator that is linear with respect to infinitesimal $\delta\phi$, defined following the concept of Fr\'echet derivative:
\begin{equation}
	G'[\phi,\delta\phi]=\delta G[\phi].
\end{equation}
It is easy to find that after the T$^1$ equation is elevated to T$^n$, the $(p,q)$ value of each non-zero term on the right-hand side originates from the summation of $n$ existing $(p,q)$ pairs (repetition allowed) from the T$^1$ equation, with the sum of $q$ reduced by $n-1$. While not observed in the cases studied here, it remains possible that some expected $(p,q)$ terms can automatically cancel out during elevation.

Next, we discuss the conditions for isotropic tensorization of the T$^n$ equation \eqref{eqn:OrderElevA}. When discussing symmetries, we use the active viewpoint, where transformations are performed on the field instead of the coordinate system.

For an equation involving an $m$-th order tensor $\Phi_{i_1\cdots i_m}$ ($m\in\mathbb{N}$, excluding pseudotensors) in $d$-dimensional Euclidean space, the matrix representing spatial rotation or reflection is denoted by $\Lambda\in\text{O}(n)\subset\mathbb{R}^{d\times d}$. Isotropy means that the equation does not depend on a specific direction and remains invariant under the action of general $\Lambda$. In other words, both sides of the equation must transform consistently under the action of $\Lambda$. Similar to the above, we only consider $(p,q)$ terms where $p$ is the total order of derivatives and $q$ is the total power of $\Phi$. A $(p,q)$ term under the action of $\Lambda$ will involve $p+qm$ factors of $\Lambda$. Since $\Lambda$ within each term can only be eliminated via $\Lambda^T\Lambda=I$, a necessary (but not sufficient) condition for the equation to satisfy isotropy is that $p+qm$ must be either all odd or all even for non-zero $(p,q)$ terms on both sides of the equation.

However, for the simplest 1D case, deriving a necessary and sufficient condition for isotropy is possible. Next, we consider the 1D equation \eqref{eqn:OrderElevA} and a general $m$, treating $\phi$ as the sole component of the $m$-th order tensor $\Phi$ in 1D. Since $\Lambda\in\text{O}(1)=\{\pm1\}$ in 1D, only the transformation rule of the equation under $\Lambda=-1$ needs to be addressed. We define the transformation operators for 1D functions:
\begin{equation}
	\mathcal{P}[\phi](x,t)=+\phi(-x,t),~\mathcal{N}[\phi](x,t)=-\phi(+x,t),
\end{equation}
which satisfy the following relations with $F_{p,q}$:
\begin{equation}\label{eqn:FPNpermute}
	F_{p,q}\mathcal{P}=(-1)^p\mathcal{P}F_{p,q},~
	F_{p,q}\mathcal{N}=(-1)^qF_{p,q}.
\end{equation}
The transformation of $\phi$ under $\Lambda=-1$ corresponds to the combined transformation $\mathcal{P}\mathcal{N}^m$. Therefore, we find that the necessary and sufficient condition for equation \eqref{eqn:OrderElevA} to satisfy $m$-th order tensor isotropy is that all non-zero $(p,q)$ terms in $F^{(n)}[\phi]$ satisfy $p+(q-1)m\in2\mathbb{N}$, which can be simplified case-by-case:
\begin{equation}\label{eqn:Tensor1Dcondition}
	\forall F^{(n)}_{p,q}\neq0:~
	\begin{cases}
		m\in2\mathbb{N}:~p\in2\mathbb{N},
		\\
		m\in2\mathbb{N}+1:~p+q\in2\mathbb{N}+1.
	\end{cases}
\end{equation}

If the 1D isotropy condition \eqref{eqn:Tensor1Dcondition} for the T$^n$ equation \eqref{eqn:OrderElevA} is satisfied for some $m$, then the equation \eqref{eqn:OrderElevA} can be extended into a covariant tensor equation by operations such as extending $\phi$ to the $m$-th order tensor $\Phi$, extending $\partial_x$ to $\nabla$-type operators, and performing necessary tensor contractions. This operation is denoted as the $\Lambda^m$ tensorization. However, the tensorization is often not unique. For example, the $\Lambda^1$ tensorization of $\phi\partial_x\phi$ could be $\Phi_i\partial_i\Phi_j$, $\Phi_j\partial_i\Phi_i$, or $\partial_j(\Phi_i\Phi_i)/2$. The significance of tensorization is that once the tensor form of the equation is determined, it can be readily generalized to 2D, 3D, or higher dimensions.

We define T$^n\Lambda^m$ extension as the operation of elevating the T$^1$ equation \eqref{eqn:original} to T$^n$ followed by $\Lambda^m$ tensorization. The necessary and sufficient condition for T$^n\Lambda^m$ extension is called the T$^n\Lambda^m$ extensibility. Based on the above analysis, the T$^n\Lambda^m$ extensibility for equation \eqref{eqn:original} means that all non-zero terms on the right-hand side of the corresponding T$^n$ equation \eqref{eqn:OrderElevA} satisfy equation \eqref{eqn:Tensor1Dcondition} for $m$.

Since elevation is the only option to eliminate anisotropy with original solutions preserved, if all T$^n\Lambda^m$ extensibilities are absent for equation \eqref{eqn:original}, we define this equation as inherently anisotropic. In other words, an inherently anisotropic equation cannot be extended to describe spatially isotropic systems without substantial modifications that would result in the loss of original solutions.

In addition to the form given in equation \eqref{eqn:original}, slightly more complex equations exist, such as the Camassa-Holm (CH) equation \citep{Camassa1993Anintegrable}:
\begin{equation}\label{eqn:CH}
	\partial_t\phi+2\kappa\partial_x\phi
	-\partial_x^2\partial_t\phi+3\phi\partial_x\phi
	=2\partial_x\phi\partial_x^2\phi+\phi\partial_x^3\phi,
\end{equation}
and the Degasperis-Procesi (DP) equation \citep{Degasperis1999AsymptoticIntegrability}. These equations can be rearranged and viewed as a T$^1$ equation where a linear and invertible spatial differential operator $S$ acts on the left-hand side:
\begin{equation}\label{eqn:Sdt}
	S[\partial_t\phi]=F[\phi],
\end{equation}
where $F$ also consists of $(p\in\mathbb{N},q\in\mathbb{N}_+)$ terms as in equation \eqref{eqn:original}. For this form, elevation can still be performed by applying $\partial_t$ on both sides:
\begin{equation}\label{eqn:SdtTn}
	n\geq2:~S[\partial^n_t\phi]=\tilde{F}^{(n)}[\phi]=\tilde{F}'^{(n-1)}[\phi,S^{-1}[F[\phi]]],
\end{equation}
where $\tilde{F}^{(1)}=F$. Note that for $n\geq2$, $\tilde{F}^{(n)}$ generally cannot be expressed using multiplication, addition, and local spatial differentiation. Consequently, the symmetry criterion for the 1D equation cannot be obtained simply by decomposing the $(p,q)$ terms, but should follow the fundamental principle that both sides of the equation transform  consistently under the $\Lambda=-1$ transformation. Furthermore, if $S$ contains only even-order spatial derivatives, it can be shown that a sufficient condition for the extended T$^n$ equation to satisfy $m$-th order tensor isotropy is that the $F^{(n)}$ obtained by iterating $F$ via equation \eqref{eqn:OrderElevB} satisfies equation \eqref{eqn:Tensor1Dcondition} without cancellation. In other words, in this case the T$^n\Lambda^m$ extensibility for equation \eqref{eqn:original} with identical $F$ implies the T$^n\Lambda^m$ extensibility of equation \eqref{eqn:Sdt}.
\section{Examples: extensibility and inherent anisotropy}
Based on the definitions and analysis methods described above, together with the assumption that no cancellation occurs, the T$^n\Lambda^m$ extensibility for several typical T$^1$ scalar equations is evaluated and summarized in table~\ref{tab:category}. Some of these equations are written in normalized or simplified forms with undetermined coefficients. Therefore, with specific coefficients, some equations may still exhibit additional extensibility due to the cancellation of coefficients after elevation.

\begin{table}\begin{center}\begin{tabular}
{	m{3.5cm}<{\centering}		m{5cm}<{\centering}		m{3.5cm}<{\centering}	}
\toprule																	
	Equation&$F[\phi]$&Extensibility	\\
\midrule																	
	Linear advection &
	$\partial_x\phi$ &
	T$^{2\mathbb{N}_+}\Lambda^{\mathbb{N}}$
	\\
	Diffusion &
	$\partial_x^2\phi$ &
	T$^{\mathbb{N}_+}\Lambda^{\mathbb{N}}$
	\\
	Linear KdV &
	$\partial_x^3\phi$ &
	T$^{2\mathbb{N}_+}\Lambda^{\mathbb{N}}$
	\\
	Advection-diffusion &
	$\partial_x\phi+\lambda\partial_x^2\phi$ &
	$-$
	\\
	Inviscid Burgers &
	$\phi\partial_x\phi$ &
	T$^{2\mathbb{N}+1}\Lambda^{2\mathbb{N}+1}$,~T$^{2\mathbb{N}_+}\Lambda^{\mathbb{N}}$
	\\
	Viscous Burgers &
	$\phi\partial_x\phi+\lambda\partial_x^2\phi$ &
	T$^{\mathbb{N}_+}\Lambda^{2\mathbb{N}+1}$
	\\
	KdV &
	$\phi\partial_x\phi+\lambda\partial_x^3\phi$ &
	T$^{2\mathbb{N}_+}\Lambda^{2\mathbb{N}}$
	\\
	mKdV &
	$\phi^2\partial_x\phi+\lambda\partial_x^3\phi$ &
	T$^{2\mathbb{N}_+}\Lambda^{\mathbb{N}}$
	\\
	KdV-Burgers &
	$\phi\partial_x\phi+\lambda_1\partial_x^2\phi+\lambda_2\partial_x^3\phi$ &
	$-$
	\\
	mKdV-Burgers &
	$\phi^2\partial_x\phi+\lambda_1\partial_x^2\phi+\lambda_2\partial_x^3\phi$ &
	$-$
	\\
	CH/DP &
	$\partial_x\phi+\lambda_1\phi\partial_x\phi+\lambda_2\partial_x\phi\partial_x^2\phi+\lambda_3\phi\partial_x^3\phi$ &
	T$^{2\mathbb{N}_+}\Lambda^{2\mathbb{N}}$
	\\
\bottomrule																	
\end{tabular}\end{center}\caption{Extensibility classification for typical T$^1$ equations.}
\label{tab:category}\end{table}

For more specific examples, we present the extensibility analysis and associated extension operations for the unsimplified Burgers, KdV, and KdV-Burgers (KdVB) equations.

The Burgers equation \citep{Bateman1915Somerecent,Burgers1948Amathematical} with viscosity $\lambda>0$ is:
\begin{equation}\label{eqn:Burgers}
	\partial_t\phi+\phi\partial_x\phi
	=\lambda\partial_x^2\phi.
\end{equation}
Analysis shows that the Burgers equation is already isotropic for $m=1$ and exhibits T$^{\mathbb{N}_+}\Lambda^{2\mathbb{N}+1}$ extensibility. In addition to the well-known T$^{1}\Lambda^{1}$ form, $\partial_t\bm{u}+\bm{u}\cdot\nabla\bm{u}=\lambda\nabla^2\bm{u}$, equation \eqref{eqn:Burgers} can also be extended to the T$^{1}\Lambda^{3}$ form, for instance:
\begin{equation}\label{eqn:BurgersT1X3}\begin{split}
	\partial_t\Phi_{ijk}+\Phi_{ijl}\partial_l\Phi_{krr}=\partial_{ll}\Phi_{ijk},
\end{split}\end{equation}
to potentially describe systems involving third-order tensors.

As another classic nonlinear wave equation, the KdV equation \citep{Korteweg1895Onthe,Miles1981TheKorteweg} in the rest frame, with reference depth $h>0$, gravitational acceleration $g>0$, and characteristic wave speed $c=\sqrt{gh}$, can be written as
\begin{equation}\label{eqn:KdV}
	\partial_t\phi
	=F_{\pm}[\phi]
	=\pm c \left(\partial_x\phi+\frac{3}{2h}\phi\partial_x\phi+\frac{h^2}{6}\partial_x^3\phi\right).
\end{equation}
Analysis shows that the KdV equation itself is anisotropic, but it exhibits T$^{2\mathbb{N}_+}\Lambda^{2\mathbb{N}}$ extensibility. Extending the KdV equation to T$^2$ yields the KdV$^2$ equation:
\begin{equation}\label{eqn:KdV2}\begin{split}
	\partial_t^2\phi
	=&F^{(2)}_+[\phi]=F^{(2)}_-[\phi]
	\\
	=&c^2\partial_x^2\left[
	\phi+\frac{3}{2h}\phi^2+\frac{h^3}{3}\partial_x^2\phi
	+\frac{3}{4h^2}\phi^3+\frac{h}{4}\partial_x^2(\phi^2)-\frac{3h}{8}(\partial_x\phi)^2
	+\frac{h^4}{36}\partial_x^4\phi
	\right].
\end{split}\end{equation}
The fact that $F^{(2)}_+=F^{(2)}_-$ illustrates the self-consistency of the T$^2$ extension operation applied to the KdV equation. By comparing equations \eqref{eqn:KdV} and \eqref{eqn:KdV2}, we find that when $\phi$ is viewed as the interface displacement, the initial condition for the KdV equation consists only of the initial waveform, whereas the initial condition for the KdV$^2$ equation also includes the initial rate of change of the waveform. This indicates that the KdV$^2$ equation not only eliminates anisotropy but also enhances physical completeness.

The similarities and differences between the KdV and KdV$^2$ equations are worth further illustration. We define the KdV$_\pm$ equation as the equation \eqref{eqn:KdV} with $F_\pm$. The travelling-wave solutions of the KdV$_\pm$ equation, after the direction is reversed, will instead become the solutions of the KdV$_\mp$ equation. In contrast, the KdV$^2$ equation inherits all solutions and conservation laws of both the KdV$_+$ and KdV$_-$ equations, allowing the coexistence of waves travelling in both directions. In addition to all previously discovered analytical KdV$_\pm$ solutions, there also exists a new pair of travelling wave solutions for KdV$^2$:
\begin{subequations}\label{eqn:KdV2sol}\begin{gather}
	\phi(x,t)=A\text{sech}^2(k(x\pm vt))+B,~k\in\mathbb{R},
	\\
	v=c h^2 k^2,~A=\frac{10}{3}h^3 k^2,~B=-\frac{2}{9}(3+5 h^2 k^2)h,
\end{gather}\end{subequations}
which may be related to physically relevant waves.

Furthermore, comparison reveals that the $\text{KdV}^2$ equation \eqref{eqn:KdV2}, when retaining only the first three terms on the right-hand side, can precisely recover the Boussinesq equation \citep{Boussinesq1871aTh,Boussinesq1871bTh}. Therefore, the KdV$^2$ equation can be viewed as a higher-order correction obtained by adding terms such as $\partial_x^2(\phi^3)$ and $\partial_x^6\phi$ to the Boussinesq equation. Notably, the $\partial_x^6\phi$ term provides linear stability that the original Boussinesq equation lacks, as shown in the dispersion relation for infinitesimal $\phi\sim e^{i(kx-\omega t)}$ with $k\in\mathbb{R}$:
\begin{equation}\begin{split}
	\text{Boussinesq}:~&\omega=\pm ck\sqrt{1-\frac{h^3}{3}k^2},
	\\
	\text{KdV$^2$}:~&\omega=\pm ck\sqrt{1-\frac{h^3}{3}k^2+\frac{h^6}{36}k^4}
	=\pm ck\left(1-\frac{h^3}{6}k^2\right).
\end{split}\end{equation}
More specifically, the linear Boussinesq equation will lead to exponential growth at $|k|>\sqrt{3}h^{-3/2}$ due to an imaginary $\omega$, while the linear KdV$^2$ equation is stable with $\omega\in\mathbb{R}$ at all $k$.

Based on the KdV$^2$ equation, we can provide a T$^2\Lambda^0$ extension of the KdV equation:
\begin{equation}\label{eqn:KdV-T2X0}\begin{split}
	\partial_t^2\phi
	=&c^2\nabla^2\left[
	\phi+\frac{3}{2h}\phi^2+\frac{h^3}{3}\nabla^2\phi
	+\frac{3}{4h^2}\phi^3+\frac{h}{4}\nabla^2(\phi^2)-\frac{3h}{8}|\nabla\phi|^2
	+\frac{h^4}{36}\nabla^4\phi
	\right].
\end{split}\end{equation}
Evidently, this form can be naturally interpreted as a valid 2D isotropic hydrodynamic wave equation. It can also be regarded as a modified 2D Boussinesq equation with additional higher-order terms providing linear stability.

Beyond the context of hydrodynamic wave dynamics where the scalar $\phi$ represents the interface perturbation height, there might be other problems mathematically analogous to the KdV equation but requiring a second-order tensor $\Phi$ to characterize the physical field of interest. In such cases, we can provide a potentially useful extension (which is not unique) of the KdV equation supported by the T$^2\Lambda^2$ extensibility:
\begin{equation}\label{eqn:KdV-T2X2}\begin{split}
	\partial_t^2\Phi
	=&c^2\nabla^2\Big[
	\Phi+\frac{3}{2h}(\Phi:\Phi)I+\frac{h^3}{3}\nabla^2\Phi
	+\frac{3}{4h^2}(\Phi:\Phi)\Phi
	\\
	&~~~~~~+\frac{h}{4}\nabla^2(\Phi\cdot\Phi)
	-\frac{3h}{8}(\nabla\cdot\Phi)(\nabla\cdot\Phi)
	+\frac{h^4}{36}\nabla^4\Phi
	\Big].
\end{split}\end{equation}

It is apparent that the Burgers and KdV equations share the same form of the nonlinear term, but the Burgers equation lacks the linear convection and third-order dispersion terms while containing a second-order dissipation term. Consequently, the KdVB equation can be regarded as the KdV equation with a dissipation term added:
\begin{equation}\label{eqn:KdVB}
	\partial_t\phi
	=\pm c \left(\partial_x\phi+\frac{3}{2h}\phi\partial_x\phi+\frac{h^2}{6}\partial_x^3\phi\right)
	+\lambda\partial_x^2\phi.
\end{equation}
Our analysis reveals that it does not exhibit any T$^n\Lambda^m$ extensibility. Therefore, the KdVB equation is inherently anisotropic, indicating some potential incompatibility between dispersion and dissipation terms from a symmetry perspective.

Furthermore, table~\ref{tab:category} reveals that the CH equation can only achieve isotropy through T$^{2\mathbb{N}_+}\Lambda^{2\mathbb{N}}$ extension. In other words, if the CH equation is to physically describe an isotropic system, its dependent variable must be a scalar or other even-order tensor; it cannot follow the original interpretation \citep{Camassa1993Anintegrable} that it denotes the fluid velocity, despite the presence of the same nonlinear term as in the Burgers equation. This demonstrates that the analysis of isotropic extensibility can also impose constraints, from a symmetry perspective, on the nature of the equation's dependent variable.
\section{Concluding remarks}
The T$^n\Lambda^m$ isotropic extension of one-dimensional and first-order-in-time scalar equations is the only solution-preserving approach to eliminate anisotropy. Exhibiting the T$^n\Lambda^m$ extensibility means that: 1) the equation, after being elevated to T$^n$, can be written as an isotropic $m$-th order tensor equation; 2) solutions of the original equation still satisfy the T$^n\Lambda^m$ extended equation; and 3) the T$^n\Lambda^m$ extended equation can be straightforwardly generalized to higher dimensions. Conversely, intrinsic anisotropy characterized by the absence of any T$^n\Lambda^m$ extensibility implies that, to describe spatially isotropic physical systems, the equation requires substantial modifications accompanied by the loss of original solutions.
\backsection[Funding]{This work was financially supported by the National Natural Science Foundation of China (Grant No. 12302284), the Ministry of Science and Technology of China (Grant No. 2024ZD0713601-06), the Ningbo Municipal Bureau of Education (Grant No. 2024A-149-G), and the Ningbo Municipal Bureau of Science and Technology (Grant Nos. 2023Z227 and 2023-DST-001).}
\backsection[Author ORCIDs]{
\\\noindent\orcidlink{0000-0001-8273-7484} Shengqi Zhang \url{https://orcid.org/0000-0001-8273-7484}.
}
\bibliographystyle{jfm}
\bibliography{Reference}

\end{document}